\documentclass[twocolumn]{webofc}
\usepackage[varg]{txfonts}   
\woctitle{?????????}
\begin{document}
\title{Acceleration of cosmic rays and gamma-ray emission from supernova remnant/molecular cloud associations}
%
%

\author{Stefano Gabici\inst{1}\fnsep\thanks{\email{gabici@apc.in2p3.fr}} \and
        Julian Krause\inst{1}\fnsep \and
        Giovanni Morlino\inst{2}\fnsep \and
        Lara Nava\inst{3}
}

\institute{APC, Univ Paris Diderot, CNRS/IN2P3, CEA/Irfu, Obs de Paris, Sorbonne Paris Cit\'e, France
\and
           INFN -- Gran Sasso Science Institute, viale F. Crispi 7, 67100 L'Aquila, Italy
\and
           Racah Institute of Physics, The Hebrew University of Jerusalem, 91904, Israel
          }

\abstract{%
The gamma-ray observations of molecular clouds associated with supernova remnants are considered one of the most promising ways to search for a solution of the problem of cosmic ray origin. Here we briefly review the status of the field, with particular emphasis on the theoretical and phenomenological aspects of the problem.
}
\maketitle
\section{Introduction}
\label{sec:intro}

In a pioneering paper, Black and Fazio \cite{blackfazio} proposed to measure the mass of Molecular Clouds (MCs) from their gamma-ray emission. This implicitly assumes the existence of an ubiquitous and uniform {\it sea} of Cosmic Rays (CRs) that pervades the whole Galaxy. Under this assumption the intensity of CRs in any given MC is known and its gamma-ray emission, dominated by the contribution from the $\pi^0$-decay channel, simply scales as $F_{\gamma} \propto M_{cl}/d^2$, where $M_{cl}$ is the mass of the cloud and $d$ its distance (see e.g. \cite{felixbook}). Thus, if the distance is also known, the mass of the cloud can be estimated by measuring $F_{\gamma}$. 

In fact, the mass of a MC can be estimated also from the intensity of the CO line emission \cite{Xco}, and thus the argument above can be reversed and one can then infer, from gamma-ray observations, the intensity of CRs into the MC under examination (see e.g. \cite{issa,felixclouds}). For this reasons MCs have been sometimes referred to as {\it cosmic ray barometers}, because they can serve as probes to estimate the CR intensity (and thus their pressure) at various locations in the Galaxy \cite{sabrinona}.

Both the aforementioned approaches have limitations. The assumption of a rough spatial homogeneity of CRs is justified only on large spatial scales, but is most likely violated in the vicinity of particle accelerators, due to the injection of freshly accelerated CRs in the surrounding interstellar medium \cite{thierry,atoyan,gabici2007}. 
On the other hand, mass estimates based on the observations of the CO line emission rely on the poorly determined CO-to-H$_2$ conversion factor, $X_{\rm CO}$, and this adds an uncertainty in the use of MC as CR barometers \cite{hartquist}. However, excesses above the CR sea of one or two order of magnitudes can be easily realized in the proximity of CR sources, clearly overwhelming the uncertainty on $X_{\rm CO}$ (see Table 1 in \cite{Xco}). To conclude, MCs can be definitely used as probes to detect the presence of large and localized enhancements in the CR intensity, and thus the presence of CR sources. Clearly, this fact has crucial implications in the quest for the origin of Galactic CRs (see \cite{gabicireview} for an extended review).

Excesses in the CR intensity have been indeed detected by means of gamma-ray observations of MCs (e.g. \cite{hessridge,hessW28,fermiW28,W44,W51}), and in most cases the MCs are located in the vicinity of, or partially overlap with SuperNova Remnants (SNRs), the putative sources of CRs. Remarkably, for two SNRs (W28 and W44) part of the gamma-ray emission comes from a region located clearly outside of the SNR shell. In this case, the observations can be interpreted as the result of the escape of CRs from the SNR shock, their propagation up to the cloud, and their interaction with the dense gas \cite{gabici2009}. In the following Section we will describe how the interpretation of these observations might lead to an indirect proof for the fact that SNRs indeed accelerate CR hadrons, and thus reinforce the belief that they might be the sources of Galactic CRs. Also, a constrain on the propagation properties of CRs in the interstellar magnetic field can be obtained from such observations.

An implicit assumption done so far is that CRs penetrate unimpeded MCs, i.e., the CR spectrum immediately outside of the MC is identical to that inside it. In fact, this might not be the case, due to the interplay of CR spatial transport at the MC bondary and energy losses suffered by CRs inside MCs \cite{skilling,morlino}. We will briefly discuss this issue in Section 3. Finally, in Section 4 we will present a multi-wavelength approach aimed at extracting the CR spectrum from MCs over more than six orders of magnitude in energy: from the sub-MeV to multi-TeV domain.

\section{Runaway cosmic rays}

\begin{figure}[t]
\centering
\includegraphics[width=.5\textwidth,clip]{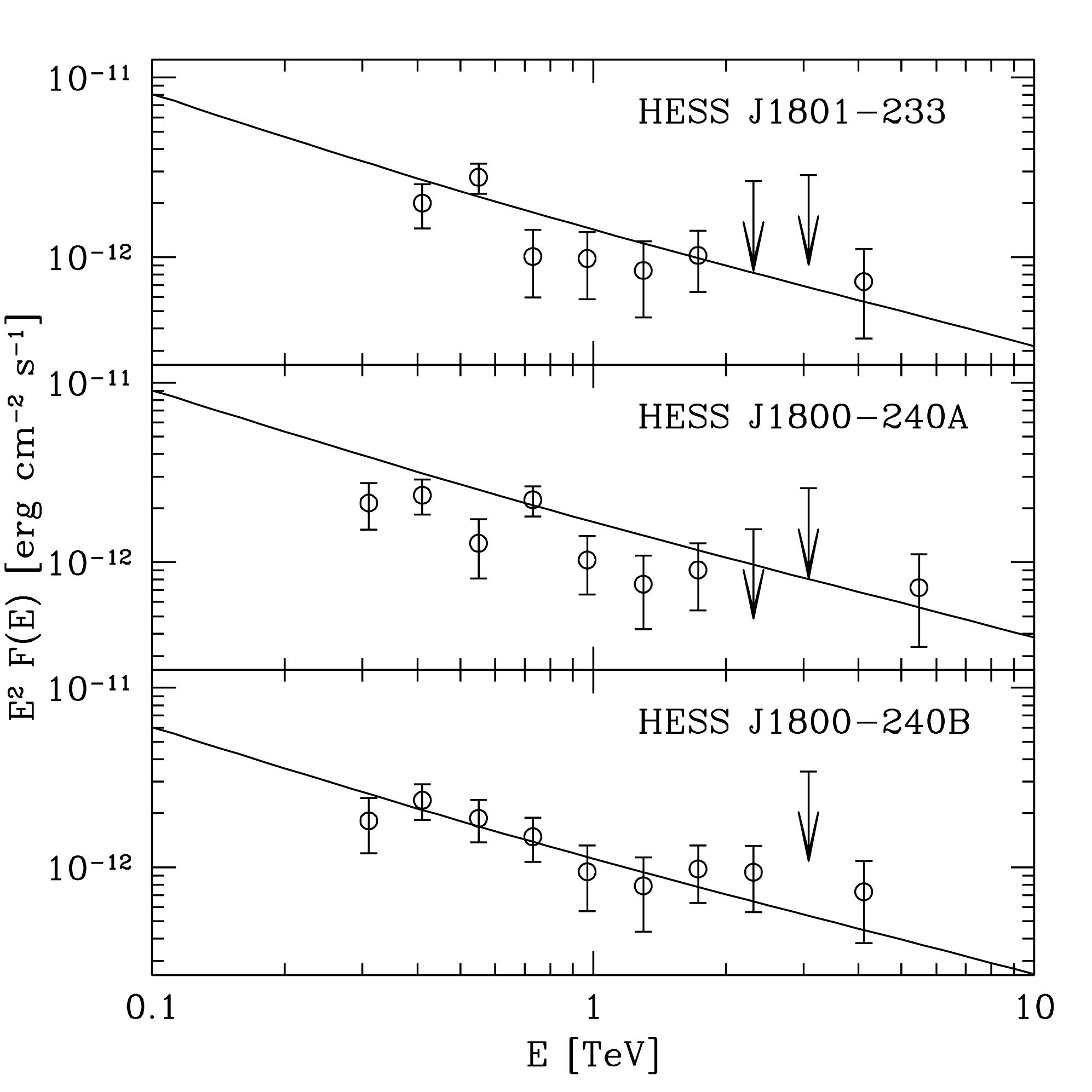}
\caption{Fit of the gamma-ray emission from three molecular clouds located in the vicinity of the supernova remnant W28. Isotropic diffusion of cosmic rays has been assumed. Data points refer to HESS observations \cite{hessW28}. Figure from \cite{gabiciW28}.}
\label{fig-1}       
\end{figure}

Though the details of the escape of CRs from SNR shocks are still debated (see e.g. \cite{bell2013}, or \cite{gabiciescape} and references therein), it is generally believed that the highest energy particles (we focus here on protons) are released first, and particles of lower energy are released gradually as the shock speed decreases. Once released, relativistic protons produce gamma-rays due to their inelastic interactions with the surrounding interstellar medium \cite{atoyan}. 

This is of particular interest if a massive MC is located in the vicinity of the CR source, given that the amount of gamma rays produced in CR interactions scales with the mass of the target material. In this context, the gamma-ray emission produced by a MC illuminated by CRs escaped from a nearby SNR has been computed in \cite{gabici2007,gabici2009} for an isotropic CR diffusion coefficient. These results have been then applied to the massive MCs located in the vicinity of the SNR W28 \cite{gabiciW28} and a good fit to the observed gamma ray emission has been presented (see Fig.~\ref{fig-1}).

The assumption of isotropy of the CR diffusion coefficient $D$ implies that within a diffusion length from the SNR, $R_d \sim \sqrt{D ~ t_{esc}}$, the CR intensity is roughly $j_{\rm CR} \propto \eta E_{\rm SN}/R_d^3$, where $E_{\rm SN} \approx 10^{51}$~erg is the supernova explosion energy and $\eta$ is an efficiency that represents the fraction of such energy converted into CRs. If SNRs are the sources of CRs it must be $\eta \approx 10 \%$.
The time since CRs left the remnant is $t_{esc}$ which, for CRs of sufficiently large energy and for sufficiently old SNRs, is comparable to the SNR age: $t_{esc} \approx t_{age}$ \cite{gabici2009}. For the case of W28 $t_{age}$ is of the order several $10^4$ years. The gamma-ray flux from MC of known mass $M_{cl}$ is then $F_{\gamma} \propto j_{\rm CR} M_{cl}/d^2$ where $d \approx 2$~kpc is the distance to W28.
If the diffusion coefficient is expressed in units of the typical galactic one $D = \chi D_{gal}$, with $D_{gal} \approx 10^{28} (E/10 ~ {\rm GeV})^{0.5}$~cm$^2$/s, one can see that the gamma-ray flux scales proportionally to the ratio $\eta/\chi^{3/2}$. Thus, with this kind of studies one can not only indirectly reveal the acceleration of hadronic CRs at SNR shocks, and possibly prove (or falsify) the SNR hypothesis for the origin of CRs, but also constrain the CR diffusion coefficient in the vicinity of the SNR. The fit in Fig.~1 has been obtained assuming $\eta/\chi^{3/2} \sim 20$, which implies that for any plausible value of the CR acceleration efficiency $\eta < 1$, the CR diffusion coefficient in the vicinity of the SNR has to be significantly smaller than the typical galactic one (i.e. $\chi \ll 1$) \cite{gabiciW28}. Similar conclusions have been independently obtained also in \cite{li,ohira,fujita}.

However, the constrains on the diffusion coefficient change significantly if the assumption of isotropy of the CR diffusion is relaxed. In fact, over scales smaller than the galactic magnetic field correlation length, the diffusion of CRs is likely to be more effective along field lines rather than perpendicular to them (see e.g. \cite{lara,malkov,giacinti}). This is shown in Fig.~2 (from \cite{lara}), where the expected CR overdensity above the galactic CR sea is mapped for a region surrounding a SNR. The supernova explosion is assumed to have happened in the centre of the plot, and an isotropic (left panel) and anisotropic (right panel) diffusion coefficient has been adopted. For an anisotropic diffusion coefficient runaway CRs remain focussed along the field lines, and thus their intensity there is correspondingly enhanced. 

\begin{figure*}
\centering
\includegraphics[width=.75\textwidth,clip]{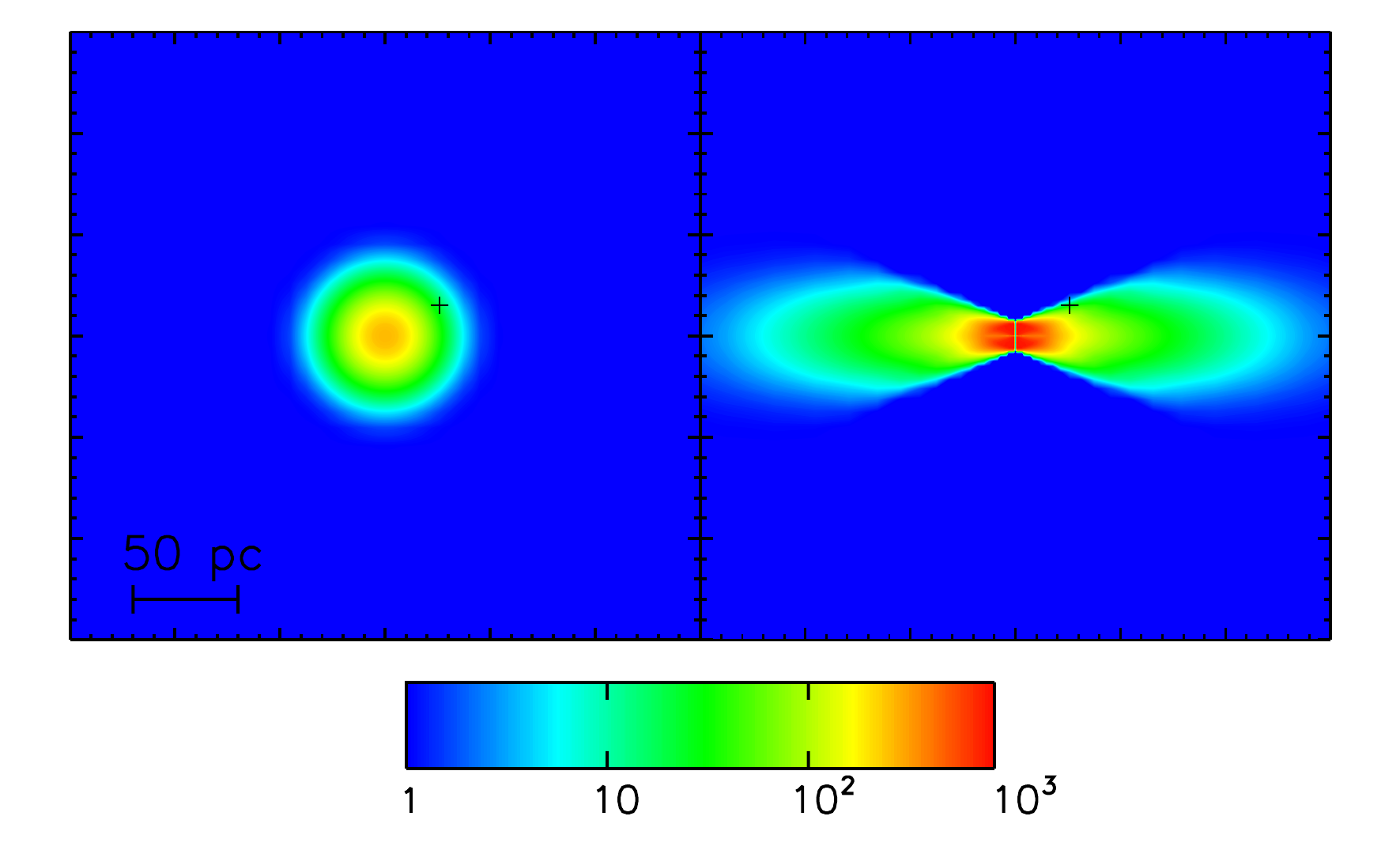}
\caption{Cosmic ray over-density around a typical supernova
remnant (see \cite{lara} for details) for a particle energy of
E = 1 TeV at a time t = 10 kyr after the explosion. The left
panel refers to an isotropic diffusion coefficient of cosmic rays
equal to $D = 5 \times 10^{26} (E/10 ~{\rm GeV})^{0.5}$~cm$^2$/s, while the right
panel refers to an anisotropic diffusion scenario with $D_{\parallel} =
10^{28} (E/10 {\rm GeV})^{0.5}$~cm$^2$/s and a magnetic field coherence length equal to 20 pc.
The black cross marks the a position at which the CR over-density
is equal in the two panels.}
\label{fig-2}       
\end{figure*}

In \cite{lara} the gamma-ray emission from the MCs in the W28 region has been fitted by assuming a fully anisotropic diffusion of CRs. To fit data, a much larger diffusion coefficient has to be assumed with respect to the case of isotropic diffusion, namely $D_{\parallel} \sim 10^{28} (E/10 ~ {\rm GeV})^{0.5}$~cm$^2$/s (which naively would correspond to $\chi \sim 1$, though a direct comparison between an isotropic and a parallel diffusion coefficient would require a more sophisticated discussion). 

Summarising, the interpretation of the gamma-ray observations of MCs illuminated by CRs accelerated at nearby SNRs dramatically depends on the assumptions made on the properties of CR diffusion. Both observations with future facilities such as the Cherenkov Telescope Array \cite{acero} and thorough theoretical investigations \cite{lara2} are needed in order to solve this issue.

\section{Penetration of cosmic rays in molecular clouds}

An assumption implicitly done so far in this paper is that of unimpeded penetration of CRs into MCs, i.e., the spectrum and intensity of CRs inside and immediately outside of the cloud are identical. However, there are reasons to believe that this might not be the case \cite{skilling,morlino,cesarsky,morfill,dogiel}. If so, the gamma-ray emission from MCs might be correspondingly affected \cite{skilling,gabici07}.

The main difficulty of studying from a theoretical point of view the problem of CR penetration in MCs resides in its strong non-linearity: CRs themselves excite the magnetic turbulence that in turn scatters them and obstructs their penetration into the cloud. The plasma instability responsible for the generation of the turbulence is the cosmic ray streaming instability \cite{kulsrud}. Such instability operates at the border of the cloud only, at the transition between ionized and neutral gas. The reason for that is that in the neutral medium inside the cloud the turbulence is immediately damped by ion-neutral friction, making the streaming instability ineffective \cite{ionneutral}.

However, it was recently pointed out in \cite{morlino} that the solution of the problem has a universal character and does not depend on the details of the plasma instability that determines the properties of CR propagation. To understand this, let us recall that the intensity of low energy CRs inside the cloud is reduced because particles undergo severe ionization energy losses there \cite{padovani}. Therefore, to match the interstellar spectrum of CRs far away from the cloud, a spatial gradient in the CR intensity must appear at the outskirts of the cloud. This is the region where CR streaming instability operates, and where CRs generate the magnetic turbulence that scatters them and determines their transport properties. The crucial result obtained in \cite{morlino} is that such a gradient that forms in the outskirts of the cloud acts as a self-regulating buffer for the flux of CRs entering the cloud. If streaming instability is very efficient (or inefficient) then the extension of the zone characterized by the gradient changes accordingly, shrinking (or expanding) and keeping in this way the flux of the CRs into the cloud constant and equal to the universal value $v_A f_0$, where $v_A$ is the Alfven speed in the diffuse interstellar medium (the velocity of the magnetic perturbations of the field) and $f_0$ the interstellar spectrum of CRs. Thus, the spectrum of CRs inside the MC can be described in terms of a characteristic energy \cite{morlino}:
\begin{equation}
E_{br} \approx 70 \left( \frac{v_A}{100 ~ {\rm km/s}} \right)^{-2/\alpha} \left( \frac{N_H}{3 \times 10^{21} {\rm cm}^{-2}} \right)^{2/\alpha} ~ \rm MeV
\end{equation}
where $N_H$ is the cloud gas column density and $\alpha \approx 2.58$ is the slope of the ionization energy loss time as a function of the particle momentum: $\tau_{ion} \propto p^{\alpha}$.
Particles of energy larger than $E_{br}$ are unaffected by energy losses, thus the CR spectrum inside and outside of the MC is identical in this energy domain. On the other hand, particles with energies smaller than $E_{br}$ suffer severe ionization losses so that the CR spectrum into the cloud is affected significantly. If the CR spectrum in the interstellar medium is a power law in momentum $f_0 \propto p^{-s}$, then the spectrum inside the cloud at particle energies smaller than $E_{br}$ reads:
\begin{eqnarray}
f_c &\propto& p^{\alpha-3}   ~~~~~~~~~ s < \alpha-3 \nonumber \\ 
f_c &\propto& p^{\alpha-s}   ~~~~~~~~~ s > \alpha-3 
\end{eqnarray}
Remarkably, for a CR spectrum in the interstellar medium $f_0(p) \sim p
^{-s}$ with $s = \alpha-3 \approx  0.42$ the slope of the spectrum
of CR is identical inside and outside of the cloud. CR spectra inside and outside of a MC of column density $3 \times 10^{21}$~cm$^{-2}$ are plotted in Fig.~3.

To conclude, we note that a break in the spectrum in the
$\approx$~100 MeV range (see expression for $E_{br}$, Eq.~1) range would not affect significantly the gamma-ray
luminosity of the cloud, because the threshold for neutral
pion production in proton-proton interactions is at a larger
energy, namely $\simeq$~280 MeV. A possibly unique exception to this is the Sgr B2 MC complex, the most massive MC complex in the Galaxy, characterised by an extremely large mass ($\approx 10^7 M_{\odot}$) and gas column density (up to $2.5 \times 10^{24}$~cm$^{-2}$ \cite{protheroe}). The gamma-ray emission from such cloud reveals a break in the CR parent spectrum at a particle energy of $\sim 4$~GeV \cite{ruigi}, which might be a manifestation of Eq.~1.

\begin{figure*}
\centering
\includegraphics[width=.75\textwidth,clip]{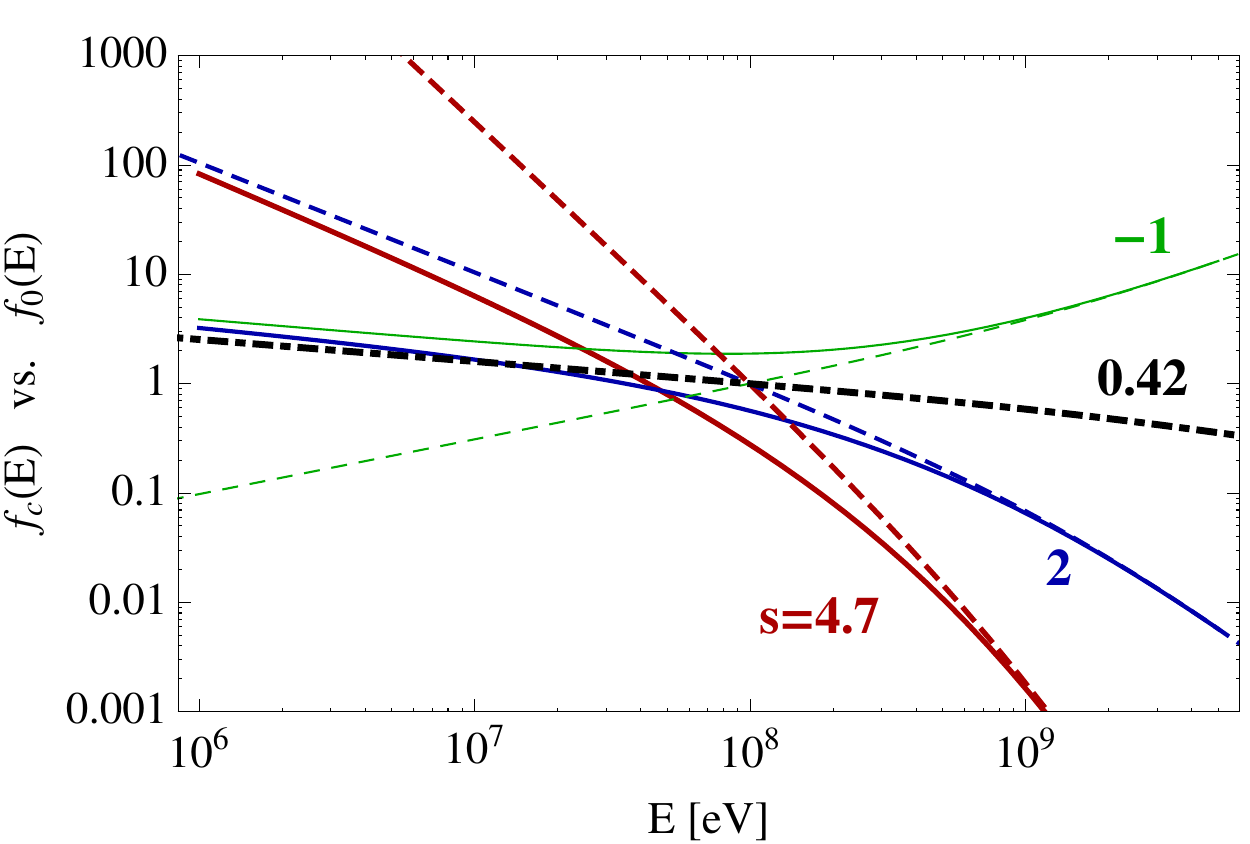}
\caption{Spectra of cosmic rays inside a molecular cloud of column density $N_H = 3 \times 10^{21}$~cm$^{-2}$ (solid lines) assuming
that the spectra far for the cloud are given by power law in momentum
of slope s (dashed lines). The dot-dashed line shows the
eigenfunction of the problem.}
\label{fig-3}       
\end{figure*}

\section{A multi-wavelength approach: from millimeter to TeV observations}

The importance of studying the amount of penetration of low energy CRs into MCs is connected to the fact that CRs are the only ionizing agents capable of penetrating large gas column densities and thus regulate the ionization fraction of dense MC cores.
The ionization fraction in turns determines the level of coupling between the magnetic field and the gas, thus playing a pivotal role in the dynamical evolution of MCs \cite{shu}.

After an early interest on this issue (see e.g. \cite{skilling,cesarsky,morfill,dogiel}, theoretical investigations of the level of penetration of low energy cosmic rays into molecular clouds became quite rare, mainly due to the difficulties connected to the measurements of both the ionization level of clouds (but see e.g. \cite{caselli} for a notable exception) and of the interstellar cosmic ray spectrum at energies low enough ($\approx$~MeV) to be relevant for gas ionization  (e.g. \cite{webber}). The breakthrough came during the last decade when observations of the infrared H$_3^+$ line and of the DCO$^+$/HCO$^+$ millimeter lines ratio from molecular clouds provided us with the measurement of the cosmic ray ionization rate from a quite significant sample of clouds (e.g. \cite{mccall}, for reviews see \cite{cecilia,indriolo}). This revived the interest in theoretical studies of the cosmic ray penetration into clouds \cite{padovani,everett}. In addition to that, the Voyager 1 probe, having reached a distance of 122 astronomical units from the Earth, reported on the measurement of the interstellar spectrum of CRs, i.e. the spectrum unaffected by solar modulation, down to energies of several MeV per nucleon (Krimigis et al. 2013, Stone et al. 2013). These measurements reduce dramatically the uncertainty on our knowledge of the CR spectrum in the interstellar medium (at least the local one), which however remains unconstrained below energies of several MeV, that are still quite relevant for ionization \cite{padovani}.

Also in this context, associations of SNRs with MCs are of prime interest. If a MC interacts with a SNR shock, one naturally expects to observe, besides an enhanced (hadronic) gamma-ray emission, also an increased CR ionisation rate. The relative enhancements would depend on the shape of the spectrum of CRs accelerated at the SNR shock itself, since $\approx$~MeV CRs are responsible for ionisation, while $\approx$~GeV to TeV CRs are responsible for the production of gamma-rays.

Over the last several years, CR ionisation rates at the level of $\gtrsim 10^{15}$~s$^{-1}$ have been reported for MCs interacting with the shocks of the SNRs IC~443, W51c, and W28 (see \cite{IC443,W51c,W28}, respectively). Since these values are well above the ionisation rate one would expect in an isolated cloud immersed in the galactic CR sea, such observations might constitute the smoking gun for the acceleration of low energy CRs at SNR shocks.
However, the relative contribution of CR protons and electrons to the ionisation rate still remains undetermined \cite{padovani}. Also the contribution to ionisation from X-ray photons from the SNR interior needs to be properly assessed (unlike UV photons, X-rays can penetrate quite large column densities of gas).

If the origin of the enhanced ionisation revealed in the above mentioned clouds will be found to have an hadronic origin, then a combination of low and high energy observations will allow to constrain the spectrum of CRs freshly injected in the interstellar medium by a SNR shock over an energy interval of unprecedented breadth: from the MeV to the multi-TeV domain.

As an example, let us now consider the MC in interaction with the shock of the SNR W28. A ionisation rate of $\gtrsim 2 \times 10^{-15}$~s$^{-1}$ has been measured there \cite{W28}. Also, gamma-rays of very likely hadronic origin have been detected by FERMI and H.E.S.S. in the GeV and TeV domian, respectively \cite{fermiW28,hessW28}. Then, let us fit the gamma-ray emission by assuming the underlying CR proton spectrum to be a power law in momentum. This requires a spectral slope of 4.8, or $4 \pi p^2 N(p) \propto p^{-2.8}$. The normalisation of $N(p)$ can be derived from the measurement of the gamma-ray flux of the MC and from the values of the MC mass and distance, which have been estimated to be equal to $5 \times 10^4 M_{\odot}$ and 2 kpc, respectively. Once the normalisation is obtained, the expected ionisation rate due to CR protons can be computed\footnote{For simplicity, we consider here a pure proton composition for CRs and a pure molecular hydrogen composition for the target material.}. This of course will depend on the extension of the CR spectrum to low energies, which are more relevant for ionisation. If the CR proton spectrum is abruptly truncated below a particle energy of 280, 100, and 60 MeV, the corresponding ionisation rates would be $\approx 4 \times 10^{-16}$, $10^{-15}$, and $2 \times 10^{-15}$~s$^{-1}$, respectively. This implies that in order to be consistent with the measured ionisation rate reported in \cite{W28}, the CR proton spectrum cannot continue with the same slope below particle energies of $\approx 50-100$~MeV. A break or a cutoff must be present in the CR spectrum. Also, it seems that there is not much room left for electron and/or X-ray induced ionisation of the gas, unless the CR proton spectrum is cutted off abruptly at a particle energy just below or of the order of the threshold for neutral pion production ($\approx 280$~MeV). 

These simple estimates exemplify how constrains on the shape of the CR spectrum can be obtained by combining low and high energy observations of MCs interacting with SNRs. This would shed new light on the process of particle acceleration at SNR shocks, and on the transport of CRs after their injection into the interstellar medium. A complete understanding of these aspects of CR physics is needed in order to verify (or not) the SNR hypothesis for the origin of galactic CRs.




{\it Acknowledgements:} SG thanks the organiser of the SUGAR2015 workshop for their invitation and financial support. He also acknowledges support from the UnivEarthS Labex program at Sorbonne Paris Cit\'e (ANR-10-LABX-0023/ANR-11-IDEX-0005-02). JK acknowledges support from the Humboldt Foundation under a Feodor Lynen fellowship. LN is supported by a Marie CurieIntra-European Fellowship of the European Community's 7th Framework Programme(PIEF-GA-2013-627715).

\end{document}